\documentclass[journal]{IEEEtran}

\usepackage{cite}

   \usepackage[dvipdf]{graphicx}
\usepackage{amsmath}
\usepackage{amssymb}

\begin{document}
\title{Experiment to Form and Characterize a Section of a Spherically Imploding Plasma Liner}



\author{S.~C.~Hsu, 
	S.~J.~Langendorf, K.~C. Yates, J. P. Dunn,
	S.~Brockington, A.~Case, E.~Cruz, F.~D.~Witherspoon,
	M.~A.~Gilmore,
	J.~T.~Cassibry, R.~Samulyak, P. Stoltz, K.~Schillo, W.~Shih, K.~Beckwith, 
	and Y.~C.~F.~Thio}
\thanks{S. C. Hsu, S. J. Langendorf, and J. P. Dunn are with Physics Division,
Los Alamos National Laboratory, Los Alamos, NM; email:  scotthsu@lanl.gov.}%
\thanks{K. C. Yates and M. A. Gilmore are with the Electrical and Computer Engineering
Department, University of New Mexico, Albuquerque, NM.}
\thanks{S. Brockington, A. Case, E. Cruz, and F. D. Witherspoon are with HyperV Technologies Corp.
and HyperJet Fusion Corporation, both in Chantilly, VA.}
\thanks{J. T. Cassibry and K. Schillo are with the Propulsion Research Center,
University of Alabama in Huntsville, Huntsville, AL.}%
\thanks{R. Samulyak and W. Shih are with the Department of Applied Mathematics and Statistics, Stony Brook University, Stony Brook, NY.}%
\thanks{P. Stoltz is and K. Beckwith was with Tech-X Corporation, Boulder, CO; K. Beckwith is now
with Sandia National Laboratories, Albuquerque, NM.}
\thanks{Y. C. F. Thio is with HyperJet Fusion Corporation, Chantilly, VA.}%

\markboth{Submitted to IEEE Trans.\ Plasma Sci.}{}
%


\IEEEspecialpapernotice{(Invited Paper)}
\maketitle
\begin{abstract}
We describe an experiment to form and characterize a section of a spherically imploding
plasma liner by merging six supersonic plasma jets that are launched by 
newly designed contoured-gap coaxial plasma guns.
This experiment is a prelude to forming a fully spherical imploding plasma liner using many dozens of
plasma guns, as a standoff driver for plasma-jet-driven magneto-inertial fusion.
The objectives of the six-jet experiments are to assess the evolution and
scalings of liner Mach number and uniformity, which are important metrics for spherically
imploding plasma liners to compress magnetized target plasmas to fusion conditions.
This paper describes the design of the coaxial plasma guns,
experimental characterization of the plasma jets,
six-jet experimental setup and diagnostics, initial diagnostic data from three- and
six-jet experiments, and the high-level objectives of associated numerical modeling.
\end{abstract}

\begin{IEEEkeywords}
Plasma applications, Plasmas
\end{IEEEkeywords}

%

\section{Introduction}
%
%
%
%
\IEEEPARstart{S}{pherically} imploding plasma liners formed by merging supersonic
plasma jets are a proposed low-cost, high-shot-rate,
standoff driver for plasma-jet-driven magneto-inertial fusion (PJMIF) \cite{thio99,hsu12ieee,knapp14}.
Magneto-inertial fusion (MIF) \cite{lindemuth83, kirkpatrick95,lindemuth09}
seeks to achieve fusion
at ion densities intermediate between those of magnetic and inertial fusion, by combining attributes of
both the latter approaches.  The primary near-term objective of the
Plasma Liner Experiment (PLX) \cite{hsu12ieee, hsu15jpp} is to demonstrate the formation of
spherically imploding plasma liners by
 merging dozens of supersonic plasma jets, and to demonstrate their viability and scalability toward
 reactor-relevant energies and scales.
 
 The PLX facility was built in 2010--2011 at Los Alamos National Laboratory (LANL) to study plasma-liner
 formation via merging supersonic plasma jets \cite{hsu09,hsu12ieee}.  From 2011--2014, PLX utilized parallel-plate mini-railguns
 \cite{witherspoon11,brockington12baps,case13,messer13}, designed and built by HyperV 
 Technologies Corp., in a series of experiments \cite{hsu15jpp}
 to study single-jet propagation \cite{hsu12pop}, two-jet oblique merging
 \cite{merritt13,merritt14}, and two-jet head-on merging \cite{moser15pop}.  These experiments,
 with the aide of associated numerical modeling \cite{loverich10jfe,merritt13}, led to the following key results of relevance
 to the physics of plasma-liner formation:
 (1)~confirmation of plasma-jet parameters, and characterization of their evolution and profiles,
 during approximately 1~m of jet propagation away from the railgun nozzle \cite{hsu12pop},
 (2)~identification and characterization of collisional-plasma-shock formation between 
 two obliquely merging plasma
 jets \cite{merritt13,merritt14}, largely consistent with hydrodynamic oblique shock theory, and 
 (3)~the role of rising mean-charge-state $\bar{Z}$ in keeping the dynamics between merging jets in a
 collisional regime (because of the $\bar{Z}^{-4}$ dependence of the counterstreaming
 ion--ion mean free path) \cite{moser15pop}.
  In parallel, additional theory and modeling efforts examined plasma-liner radial convergence and
 scalings in one dimension (1D) \cite{cassibry09,samulyak10,awe11,davis12,kim12}, and the effects of discrete jet merging on plasma-liner convergence in 3D \cite{cassibry12,cassibry13,kim13}.

\begin{figure}[!tb]
\centering
\includegraphics[width=3in]{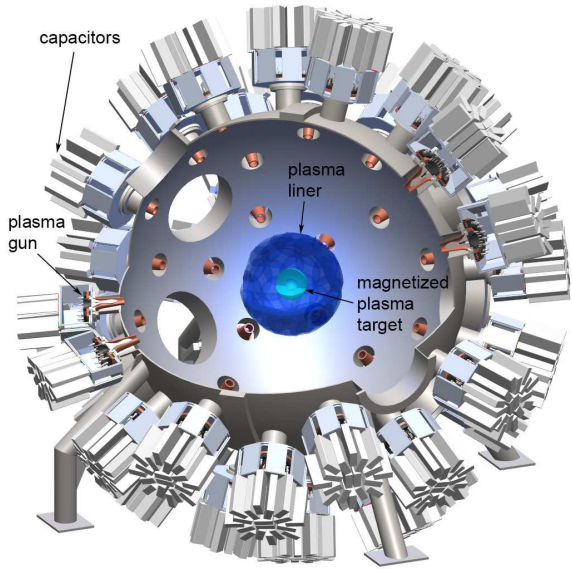}
\caption{Illustration of the PLX-$\alpha$ experimental setup, which will ultimately 
have at least 36 and up to 60 (as shown)
 coaxial plasma guns mounted around a 2.74-m-diameter vacuum chamber.  Shown are plasma
 guns with integrated capacitor banks and a spherically imploding plasma
 liner compressing a magnetized plasma target.  The PLX-$\alpha$ project is focused on developing 
 the plasma liner and is not addressing the plasma target.}
\label{fig:PLX-alpha}
\end{figure}

Collectively, these prior studies set the stage for the present PLX-$\alpha$ project, which is
named after the Accelerating Low-cost Plasma Heating and Assembly (ALPHA) program of
the Advanced Research Projects Agency--Energy (ARPA-E) that sponsors the ongoing research.
The primary objective of PLX-$\alpha$ by the end of the ALPHA program is
to form and study a fully spherical imploding plasma liner with at least 36 and up to 60 merging
plasma jets (Fig.~\ref{fig:PLX-alpha})
that are launched by newly designed coaxial plasma guns fabricated by
HyperV Technologies Corp., which is now owned by new fusion startup HyperJet Fusion Corporation.
Development of novel, $\beta>1$ (where $\beta$ is the ratio of thermal-to-magnetic pressure)
magnetized plasma targets \cite{ryutov09} for PJMIF is at a nascent stage and discussed
elsewhere~\cite{hsu12ieee,welch12,welch14}.  In future experiments involving plasma-liner 
compression of magnetized plasma targets, we envision using a subset of
the same coaxial guns (that form the liner) to form an inertially confined dense plasma target 
($\sim 10^{18}$~cm$^{-3}$),
and to potentially magnetize the target ($\beta > 1$, $\omega \tau \gtrsim 1$,
where $\omega \tau$ is the Hall magnetization parameter) using laser beat-wave 
current drive \cite{welch12,welch14}.

The remainder of this paper is organized as follows.  Section~\ref{sec:guns} describes the
design of the new PLX-$\alpha$ coaxial plasma guns and experimental data characterizing
the plasma jets that they launch.  Section~\ref{sec:six_gun} describes the motivation,
experimental/diagnostic setups, and 
initial diagnostic results of three- and six-gun experiments (a predecessor to the spherical
plasma-liner-formation experiments using 36--60 guns).  Section~\ref{sec:modeling}
provides a brief overview of the PLX-$\alpha$ numerical-modeling objectives.  Finally, Sec.~\ref{sec:summary} provides a summary and
description of future plans.

\section{PLX-$\alpha$ coaxial plasma guns}
\label{sec:guns}

New contoured-gap coaxial plasma guns
were designed and fabricated for the PLX-$\alpha$ project.  The rationale for
the contoured-gap coaxial-gun concept with pre-ionized mass injection
for launching high-mass, high-density jets to $>50$~km/s
was previously laid out in a series of seminal works \cite{thio02,cassibry06,witherspoon09}.
The design of the new PLX-$\alpha$ coaxial guns was governed by the need to achieve
particular
plasma-jet performance parameters to meet requirements of the PLX-$\alpha$ project (see
Sec.~\ref{sec:jet_requirements}), while establishing a basis for
a gun design that could be further developed and scaled up to become fusion relevant.

\subsection{Plasma-jet requirements}
\label{sec:jet_requirements}

Plasma-jet requirements were determined largely based on the desire to
build the lowest-cost experiment
that would allow studies of plasma-liner formation and convergence in reactor-relevant physics limits
inferred from the parameter regimes studied in \cite{knapp14}.  These limits are:
(1)~plasma-jet merging occurs in the collisional limit, i.e.,
the jet interpenetration depth is small compared to the jet radius, (2)~the plasma equation-of-state
(EOS)
has sufficient ionization and excitation states to provide a significant energy sink
(including strong radiative losses) compared to the
thermal energy of the jet, and (3)~the plasma flow is strongly supersonic, i.e., the sonic Mach number
$M\equiv V_{\rm jet}/C_s \gtrsim 10$, where $V_{\rm jet}$ is the directed jet speed
and $C_s$ the jet internal sound speed.
Requirements (1) and (3) lead to minimum allowable jet density and velocity, respectively.  Requirements (2) and (3) necessitate the use of heavier species such
as Ar, Kr, or Xe, though we use lighter elements as well for establishing a scaling database.
Table~\ref{table:jet_parameters} summarizes the required and
achieved plasma-jet parameters of the PLX-$\alpha$ guns.  Further discussion of the
achieved plasma-jet parameters is presented in Sec.~\ref{sec:jet_characterization}.

\begin{table}[!tb]
\renewcommand{\arraystretch}{1.2}
\caption{Required and achieved argon plasma-jet parameters of the new PLX-$\alpha$ coaxial
plasma guns.  Details regarding the measurement of jet parameters are given in
Sec.~\ref{sec:jet_characterization}.}
\label{table:jet_parameters}
\centering
\begin{tabular}{lcc}
\hline\hline
Parameter & Required & Achieved \\
\hline
Density & $\approx 2\times 10^{16}$~cm$^{-3}$ & $>2\times 10^{16}$~cm$^{-3}$\\
Mass & $>1$~mg & $0.47 \pm 0.11$~mg\\
Velocity & $\ge 50$~km/s & $52.5\pm 5.1$~km/s\\
Length & $\le 10$~cm & $16\pm 5.7$~cm\\
\hline\hline
\end{tabular}
\end{table}

\subsection{Coaxial-gun design}
\label{sec:gun_design}

To fulfill these requirements, we exploited prior HyperV gun-development efforts that had
already led to (1)~linear railguns capable of achieving the plasma-jet parameters with regard
to mass, density, and velocity~\cite{brockington12baps,hsu15jpp}
and (2)~coaxial guns with much lower current density that
used ablative mass injection~\cite{witherspoon09}.  The new PLX-$\alpha$ guns essentially combine
the plasma-jet
performance and gaseous injection of the railguns with the coaxial electrode geometry of
the coaxial guns.  

\begin{figure}[!b]
\centering
\includegraphics[width=3in]{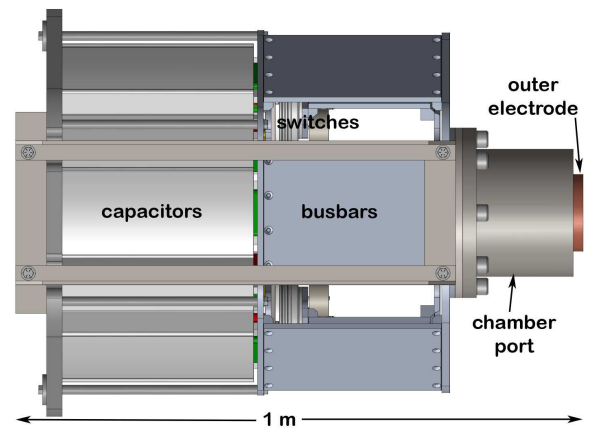}
\caption{Illustration of the PLX-$\alpha$ coaxial gun with integrated pulsed-power
components.  See also Fig.~\ref{fig:PD_setup}.}
\label{fig:gun}
\end{figure}

Figure~\ref{fig:gun} shows a full-assembly drawing of the new PLX-$\alpha$ coaxial gun and its
integrated pulsed-power module.  We chose to mount the capacitors that drive the
main gun-electrode discharge onto the back of each gun both to minimize inductance and to
eliminate the complexity and cost of requiring many parallel transmission lines.
Details of the electrodes (on the right side of the figure, hidden from view)
are proprietary information of HyperJet Fusion Corporation.
The contours on both the outer and inner electrode surfaces were designed
based on a series of MACH2 \cite{peterkin98}
simulations that led to the desired, calculated plasma-jet parameters.
The primary function of the contours is to eliminate the ``blowby instability''~\cite{cassibry06}, 
which must be avoided in order to accelerate high-mass jets.

The guns each have a fast gas valve (GV) mounted at the rear end of the inner electrode to minimize
gas travel distance from the GV to the breech (i.e., the coaxial gap at the rear end of the gun) and
also twenty tungsten ``ignitor pins'' (i.e., pre-ionizers, or PI) distributed uniformly in the azimuthal 
direction around the breech.  The GVs are typically 
pressurized at up to 20~psig (for the three- and six-gun experiments described in
Sec.~\ref{sec:six_gun}) using any gas or mixture
available in a compressed-gas bottle (Ar has been our primary working gas, but we also 
use H$_2$, He, N$_2$, Ne, Kr, and Xe on PLX)\@.  Details of the 
GV and PI designs are proprietary information of HyperJet Fusion Corporation.

The gun firing sequence is as follows.  All the capacitor banks are first charged to the
desired voltages (see Secs.~\ref{sec:jet_characterization} and \ref{sec:setup} for typical values).
The GV is triggered first at $t\approx -600$~$\mu$s (for
Ar) to fill the gun breech with neutral gas.  Then the PI system is fired at
$t\approx -20$~$\mu$s to ionize the gas fill, and finally the main gun bank is triggered
(defined to be $t=0$) to accelerate the ionized plasma out of the coaxial gun.
Gun performance can be tuned by varying the GV fill pressure, capacitor-bank
voltages, and GV and PI trigger times.

\subsection{Plasma-jet characterization}
\label{sec:jet_characterization}

Experimental studies of single-plasma-jet performance
were performed at HyperV Technologies in order to optimize jet performance, and to
make progress toward meeting the requirements given in
Table~\ref{table:jet_parameters}\@.  Jet reproducibility was assessed
and the bank voltage and trigger times optimized over a 368-shot campaign firing a single coaxial
gun using argon.  During this campaign,
the gun fired correctly at the triggered time with better than 97.5\% reliability.

\begin{figure}[!b]
\centering
\includegraphics[width=3in]{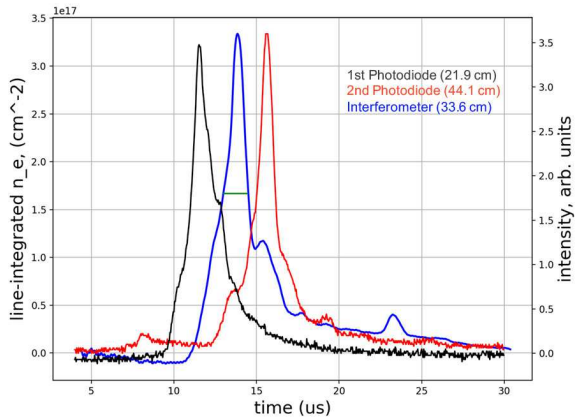}
\caption{Interferometer (cm$^{-2}$) and photodiode (arb.)\ signals vs.\ time (HyperV shot 201704251255)
for transverse views across the plasma
jet at various distances (as indicated in the legend) from the exit of the gun.  The HyperV
interferometer is a quadrature, heterodyne system using a 623-nm HeNe laser modulated at
 110~MHz \cite{case10}, and the photodiode array is similar to the system described 
 in Sec.~\ref{sec:setup}.}
\label{fig:single_jet_data}
\end{figure}

Figure~\ref{fig:single_jet_data} shows representative photodiode and interferometry data
from transverse views at various distances downstream from the exit of the gun.
These measurements were used to infer jet velocity, length, peak density, and mass as follows.
The time delay between the photodiode signals gives the jet velocity (53~km/s).  The
full-width, half-maximum (FWHM) of the interferometer signal, along with the velocity, 
gives the jet length (7.8~cm).
Assuming a jet diameter of 15~cm at the viewing position (33.6~cm downstream
of the exit of the gun), the peak electron
density $n_e$ at that position ($2.2 \times 10^{16}$~cm$^{-3}$)
is estimated by dividing the peak line-integrated electron density
$\langle n_e \ell\rangle \approx 3.3\times 10^{17}$~cm$^{-2}$ by the diameter, and integration of the 
interferometer signal over the FWHM gives the FWHM jet mass (0.8~mg).  The peak $n_e$
at the gun exit would be much higher (prior to any spreading of the jet).  
Table~\ref{table:jet_parameters} shows
the averaged parameters and standard
deviations achieved over fifteen shots with the following settings:  
$V_{\rm gun}=-4.7$~kV (center gun electrode charged
negatively), $V_{\rm GV}=8.4$~kV, $V_{\rm PI}=28$~kV,
$t_{\rm GV}= -600$~$\mu$s, and 10-psig-Ar GV pressure.

As seen in Table~\ref{table:jet_parameters}, the averaged
jet performance is close to the requirements but a little shy for jet mass and length.
An upgraded
GV (rev.~10), now being tested at HyperV,
allows much higher gas-fill pressure ($>50$~psig), which is expected
to help satisfy both the jet mass and length requirements in future experiments.
The new GV is also intended to improve jet-to-jet balance as discussed in 
Sec.~\ref{sec:interferometry}.

\section{PLX-$\alpha$ three- and six-gun experiments}
\label{sec:six_gun}

The first phase of the PLX-$\alpha$ project, in addition to coaxial-gun development and testing
at HyperV Technologies,
also includes six-gun experiments at LANL (Fig.~\ref{fig:6_guns}) in order to:
\begin{figure}[!b]
\centering
\includegraphics[width=3in]{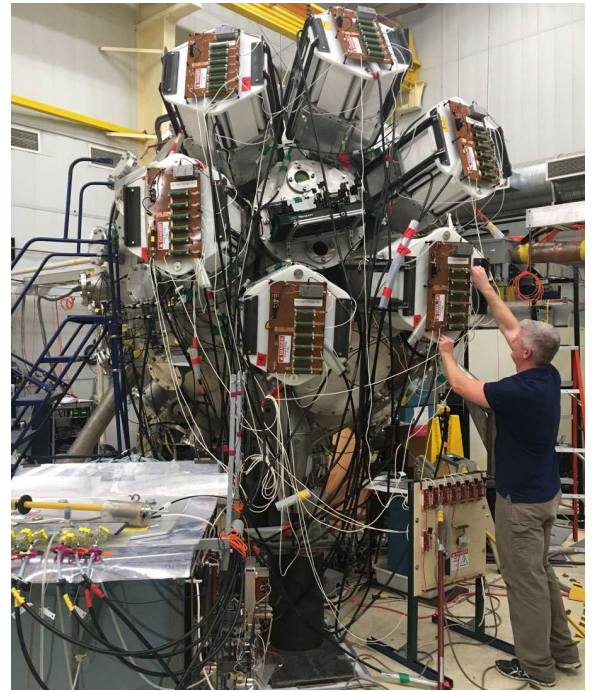}
\caption{Photograph (taken Jan.~31, 2017) of the PLX facility at LANL
with six PLX-$\alpha$ coaxial guns mounted in a hexagonal pattern
on a 2.74-m spherical vacuum chamber.  The large port in the middle of the hexagon
provides the launch positions of seven chords of a fiber-coupled, visible interferometer 
[see also Fig.~\ref{fig:int_spect_setup}(a)].  The cart at the lower left holds the capacitor
banks that power the GV and PI systems for all six guns.}
\label{fig:6_guns}
\end{figure}
(1)~successfully operate six guns simultaneously as a demonstration of technical readiness to perform
experiments using 36--60 guns for spherical plasma-liner formation, and (2)~assess the effects of
discrete jet merging on plasma-liner formation that could
cause significant degradations from one-dimensional (1D) imploding-plasma-liner
performance~\cite{knapp14,langendorf17pop}.  These jet-merging
effects include $M$-degradation due to shock heating and seeding of non-uniformities
that could exacerbate deceleration-phase instabilities at the liner/target interface (in
future, integrated liner-on-target experiments).  The
instabilities could lead to liner/target mix and reduce the overall effectiveness of plasma liners as a compression driver of magnetized plasma targets to fusion conditions.

\subsection{Experimental and diagnostic setups}
\label{sec:setup}

Figure~\ref{fig:6_guns} shows the six-gun configuration, i.e., a hexagonal arrangement
with $24^\circ$ between adjacent guns.  We can fire all six guns
simultaneously or any arbitrary subset of them.  We often fired two or three guns
for better diagnostic access to and interpretation of plasma-shock evolution between adjacent 
merging jets.

Table~\ref{table:cap_banks} summarizes the specifications of the
capacitor banks driving the six plasma guns.  Each gun
has an integrated capacitor bank (-5~kV, 575~$\mu$F, 7.2~kJ)
driving the main electrode discharge. A separate capacitor 
bank (12~kV, 96~$\mu$F, 6.9~kJ) drives all six GVs of the six guns, and yet another separate capacitor 
bank (30~kV, 12~$\mu$F, 5.4~kJ) drives all six PI systems.  A final capacitor bank (-30~kV, 6~$\mu$F, 
2.7~kJ) drives the master-trigger (MT) system for all six guns.
Typical operation on PLX has utilized -4.5~kV, 8.5~kV, 24~kV, and -28~kV for the
gun-electrode, GV, PI, and MT banks, respectively. 

\begin{table}[!b]
\renewcommand{\arraystretch}{1.2}
\caption{Capacitor-bank specifications for the six-gun PLX-$\alpha$ experiments at LANL\@.}
\label{table:cap_banks}
\centering
\begin{tabular}{lccc}
\hline\hline
Banks & Capacitor & Total bank  & Rated bank \\
 & (\#/bank) & capacitance ($\mu$F) & voltage (kV) \\
\hline
Gun & NWL 13339 & 575 & -5 \\
electrodes & (18) & per gun & \\
\hline
Gas & Maxwell 32567 & 96 & 12\\
valve (GV)  & (4) & for 6 guns & \\
\hline
Pre- & Maxwell 32814 &12 & 30 \\
ionizers (PI) & (2) & for 6 guns & \\
\hline
Master & Maxwell 32814 & 6 & -30 \\
trigger (MT) & (1) & for 6 guns & \\
\hline\hline
\end{tabular}
\end{table}

\begin{figure}[!tb]
\centering
\includegraphics[width=3in]{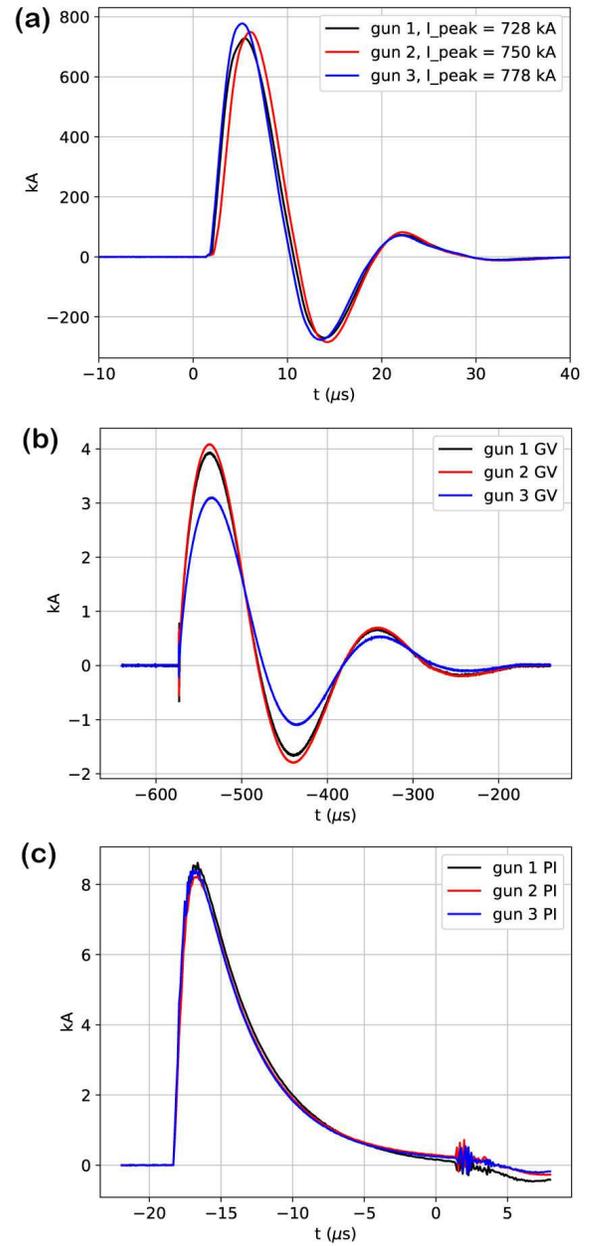}
\caption{Electrical-current waveforms, measured by Rogowski coils,
for (a)~gun electrodes, (b)~GVs, and (c)~PIs for
a representative three-gun experiment (shot 1062; charge voltages: 5~kV
for the gun electrodes, 10~kV for the GVs, and -24~kV for the PIs).  These waveforms
were absolutely calibrated to units of kA by equating the time-integrated value of each signal with
the known, stored charge in their respective capacitor banks.}
\label{fig:currents}
\end{figure}

All banks are switched by
high-voltage, high-current spark-gap switches custom made by HyperV Technologies.  Each 
575-$\mu$F bank
driving the gun electrodes consists of six separate sub-banks driven by six separate spark-gap
switches.  These
switches (6 per gun, and 36 total for 6 guns) are triggered by the MT bank.  The switches are 
pressurized to various static pressures with either Ar or Ar/N$_2$ (90\%/10\%) mixture
(gun switches) or N$_2$ (GV, PI, and MT 
switches); the switch gases are purged for 15--30~s after each experimental shot.
All switches are triggered by optically coupled signals that largely eliminate noise-induced misfires.
Figure~\ref{fig:currents} shows sample electrical currents from the gun, GV, and PI
banks for a three-gun experiment.  Note the large imbalance for one of the GV currents; this is
consistent with observed jet-to-jet imbalance leading to ongoing improvements
in the GV design and power delivery, as discussed in Sec.~\ref{sec:interferometry}.

\begin{figure}[!tb]
\centering
\includegraphics[width=3truein]{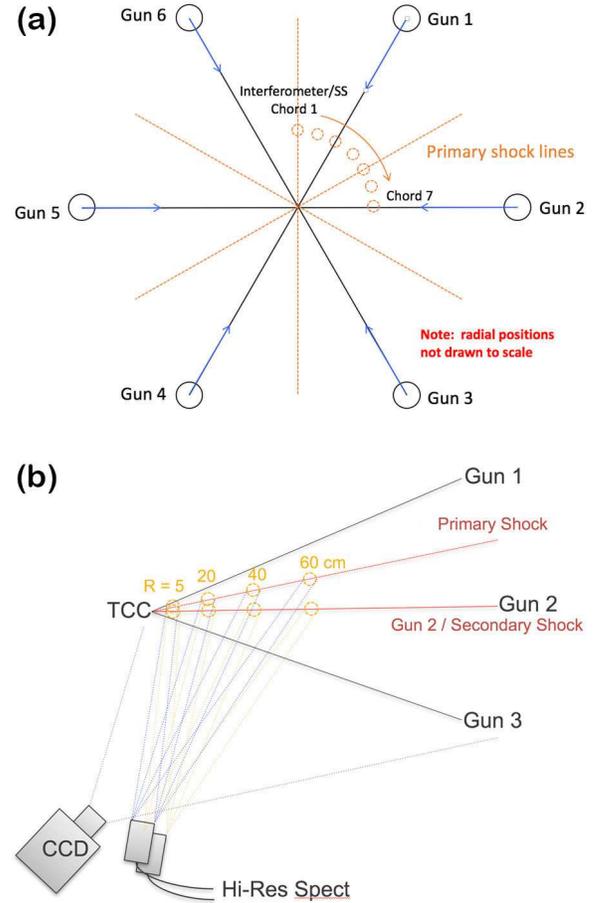}
\caption{(a)~Projected view of interferometer chords 1--7 with respect to
the location of the six guns as viewed from behind the guns.  The chords go straight across the
chamber maintaining parallel separation, whereas the guns point toward target chamber center (TCC)\@.  The survey-spectrometer collection
optic is moved around to overlap with any of the seven interferometer chords.  In the projected
plane (as depicted), the arc formed by the interferometer chords is 5.7~cm from TCC\@.
However, where chords 3 and 5 actually
intersect the jet-propagation axes is 14.7~cm from TCC\@. (b)~CCD-camera (single frame) and Doppler-spectroscopy viewing positions 
(side-on view of the jets propagating toward TCC\@).  The two 
Doppler-spectroscopy chords can be separately moved around to the various radii shown along the
``primary-shock" line or the ``secondary-shock" line.}
\label{fig:int_spect_setup}
\end{figure}

\begin{figure}[!tb]
\centering
\includegraphics[width=2.2truein]{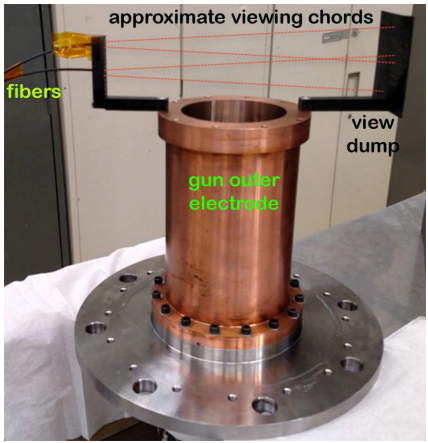}
\caption{Photograph showing the two photodiode views
that provide precise measurements of the plasma-jet velocity just beyond
the end of the gun nozzle.  The photodiode mounting structures
and view dump are made from Noryl, chosen for its electrical-insulation and low-outgassing
properties.}
\label{fig:PD_setup}
\end{figure}

Diagnostics for the six-gun experiments are summarized in Table~\ref{table:diagnostics}\@.
\begin{table}[!tb]
\renewcommand{\arraystretch}{1.2}
\caption{Summary of diagnostics for the six-jet experiments.}
\label{table:diagnostics}
\centering
\begin{tabular}{lcc}
\hline\hline
Diagnostic & Measurement & Parameter(s) inferred \\
\hline
12-chord & phase shift & line-integrated electron\\
interferometer & (time resolved) & density $\langle n_e\ell\rangle$ \\
\hline
survey & visible line spectra & chord-averaged\\
spectrometer & (time gated) & $n_e$, $T_e$, $Z$\\
\hline
high-resolution & visible line spectra & line-integrated \\
spectrometer & (time gated) & $T_i$ and Doppler shift\\
\hline
photodiode & visible light & $V_{\rm jet}$ and\\
array & (time resolved) & axial profile\\
\hline
CCD camera & visible image & $V_{\rm jet}$ and\\
(single frame) & (time gated) & global plasma appearance\\
\hline
CCD camera & visible image sequence & same as above\\
(12 frames) & (time gated) & and plasma evolution\\
\hline\hline
\end{tabular}
\end{table}
The twelve-chord, fiber-coupled, visible interferometer (using a 320-mW, 651-nm
solid-state laser and upgraded from a previous
eight-chord system \cite{merritt12a,merritt12b}) and broadband visible survey spectrometer
have both been described in detail elsewhere \cite{hsu12pop, hsu15jpp}.  The
survey-spectrometer detector (0.160~nm/pixel resolution at 510~nm)
is now upgraded to a PI-MAX2 intensified charge-coupled-device
(CCD) camera ($1024\times256$~pixels, 16-bit dynamic range, typical exposure of 1--2~$\mu$s),
and the collection
optic has also been upgraded such that the diameter of the viewing chord at the positions
of interest within the plasma is about 1~cm.  Their setups (initially,
using only seven of the twelve interferometer chords) are shown in Fig.~\ref{fig:int_spect_setup}(a).
The high-resolution spectrometer is a 4-m McPherson 2062DP, with 2400~mm$^{-1}$ grating
(1.52~pm/pixel at 480.6~nm),
coupled to a Stanford Computer Optics 4 Quik~E intensified-CCD camera
($752\times 482$~pixels, 10-bit dynamic range, typical exposure of 1~$\mu$s); plasma light is
collected at a chamber window with two 2-in., 100-mm achromatic lenses 
[Fig.~\ref{fig:int_spect_setup}(b)] and transported
to the spectrometer with a bifurcated $80\times100$-$\mu$m-core fiber bundle (so that
two views can be recorded simultaneously).  The diameter of the viewing chords at the
positions of interest are about 1.5~cm.
For the photodiode arrays, two channels of light are collected through 1-mm, 5/16-in.-deep pinholes 
near the end of each gun nozzle (Fig.~\ref{fig:PD_setup}),
and transported through optical fibers (SH-4001) to a photodiode-array board that
digitizes the signals at 100~MHz with 14-bit dynamic range.  The light level received
at the photodiode board
can be attenuated by physically adjusting the distance between a gap in each fiber.
The single-frame intensified-CCD camera is a DiCam Pro ICCD ($1280\times 
1024$~pixels with 12-bit dynamic range), positioned as shown in Fig.~\ref{fig:int_spect_setup}(b),
and the 12-frame intensified-CCD camera is an Invisible Vision UHSi 12/24
($1000\times 860$~pixels with 12-bit dynamic range).

\subsection{Initial diagnostic results}
\label{sec:results}

In this subsection, we present initial, sample results from key diagnostics from a series
of three- and six-gun experiments; all shots reported here used argon.
These were our first full experimental campaigns,
and thus jet performance was not yet optimized.  Furthermore,
we operated well below peak powers/energies of our capacitor banks, as we were exploring
parameter space and did not wish to push the limits of our systems yet.

\subsubsection{CCD-camera images}
\begin{figure*}[!t]
\centering
\includegraphics[width=6.3truein]{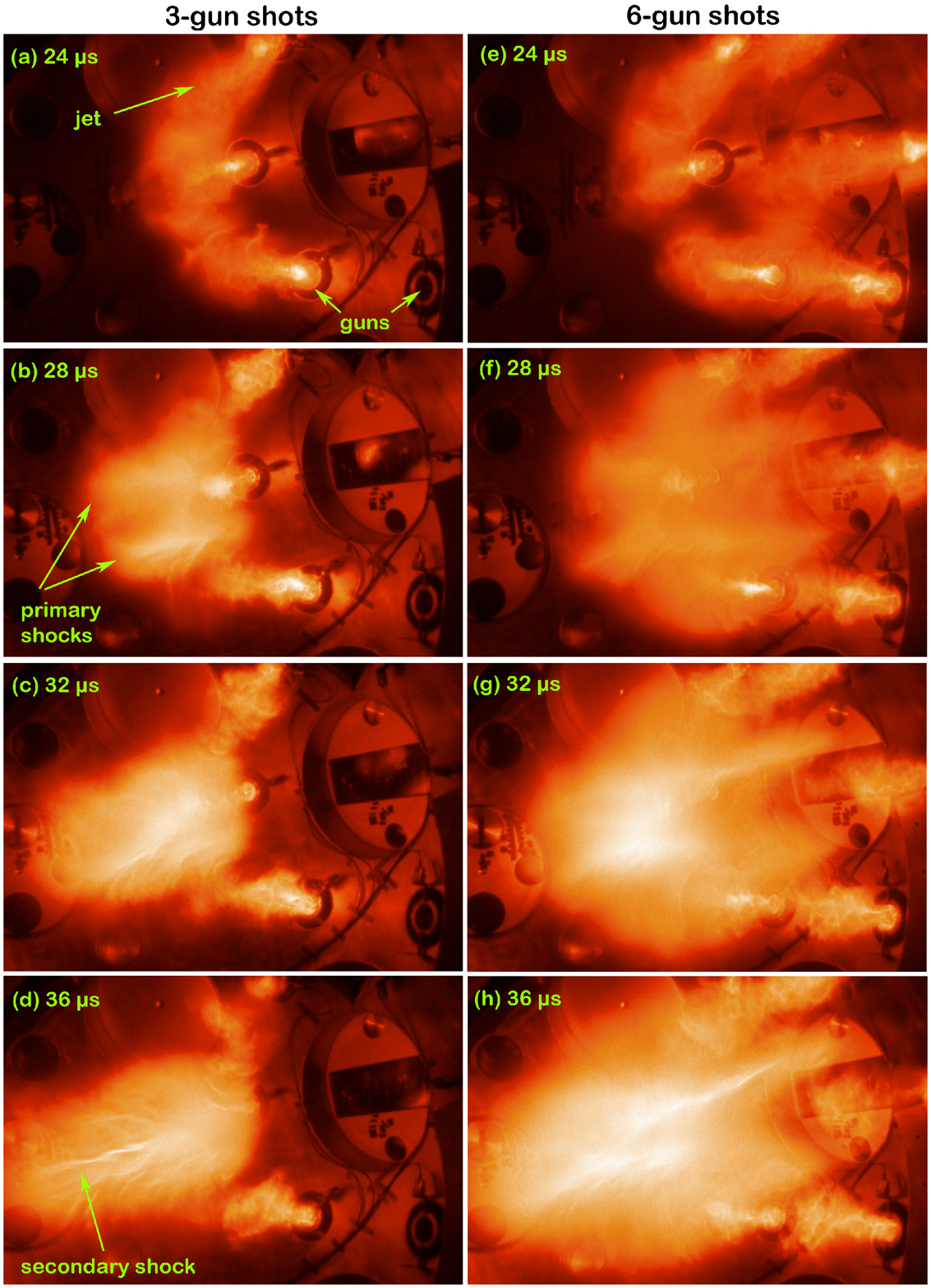}
\caption{Intensified-CCD camera images (10-ns exposure, logarithm of intensity,
false color, cropped to $1280\times 850$~pixels each) showing the evolution of
(a)--(d)~3-gun (shots 1064, 1066, 1061, 1069) and (e)--(h)~6-gun experiments
(shots 1007, 1038, 1041, 1043).  As labelled in the 3-gun image sequence,
primary shocks (b) form along the merge plane of
adjacent jets, and (presumed)
secondary shocks (d) form due to subsequent merging of the primary-shock plasmas.}
\label{fig:CCD_sequence}
\end{figure*}

Figure~\ref{fig:CCD_sequence} shows time 
series of intensified-CCD camera images (from the single-frame DiCam Pro)
from three- and six-gun experiments, showing the formation of ``primary" shocks 
(due to merging of adjacent jets)
and presumed ``secondary" shocks
(due to subsequent merging of the initial merged, shocked plasmas).
Detailed prior diagnostic studies \cite{merritt13,merritt14}  of plasma jets with similar densities and
velocities showed,
through quantitative diagnostic measurements and analyses, that what we are calling
primary shocks is consistent with collisional oblique shocks forming along the merge plane
of adjacent jets.  Primary and secondary shocks were both also observed and
studied in 3D hydrodynamic simulations \cite{kim13}.

\subsubsection{Photodiode arrays}
\begin{figure*}[!tb]
\centering
\includegraphics[width=6truein]{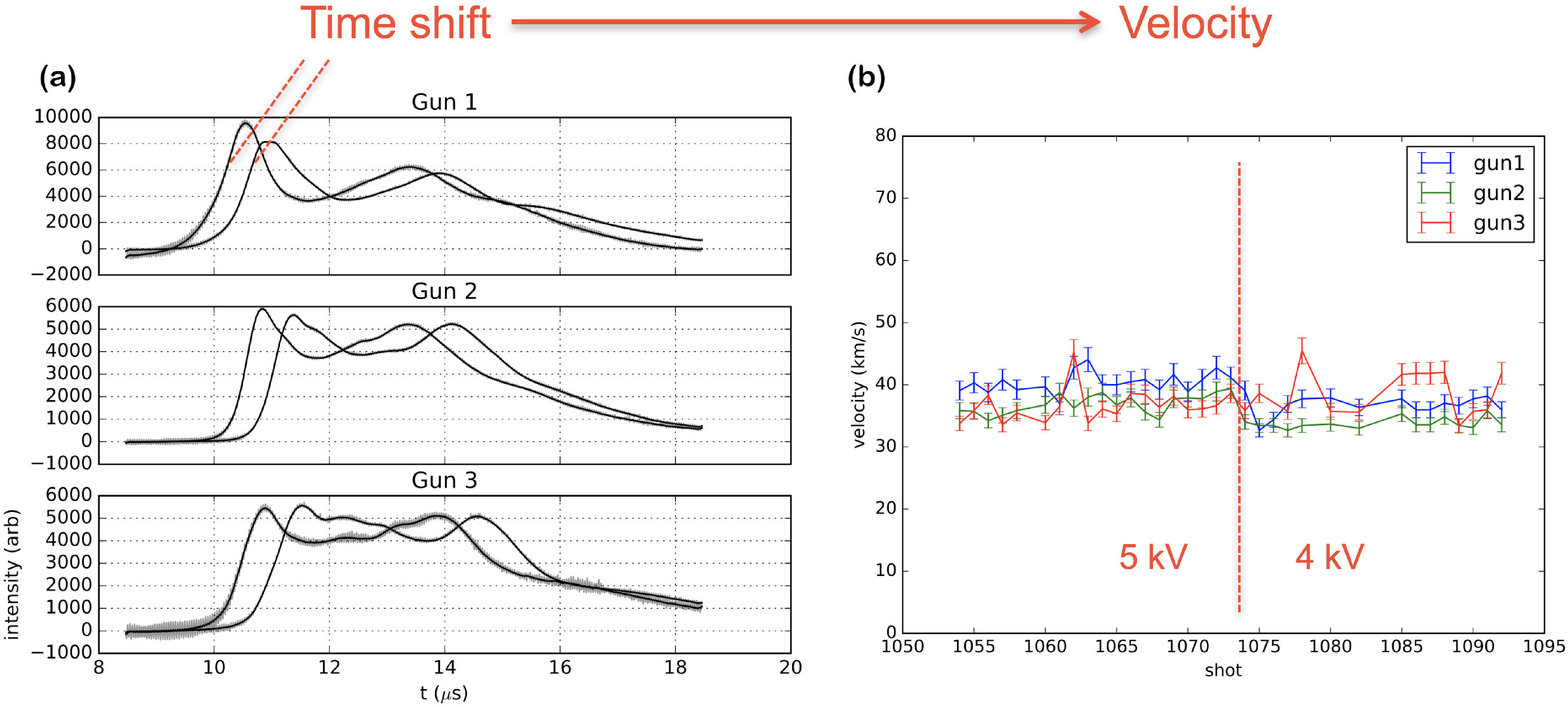}
\caption{(a)~Sample photodiode signals vs.\ time from three guns.  For each gun, photodiode viewing
chords are spaced 2~cm apart and oriented transverse to the direction of jet propagation
(see Fig.~\ref{fig:PD_setup}).
(b)~Inferred $V_{\rm jet}$ for an ensemble of shots at two gun charge voltages (4 and 5~kV).}
\label{fig:PD}
\end{figure*}

These provide measurements of $V_{\rm jet}$.
Light is collected as shown in Fig.~\ref{fig:PD_setup}; a view dump eliminates pickup of stray light and 
reflections.  Figure~\ref{fig:PD} shows
sample photodiode and the inferred $V_{\rm jet}$ values.  The latter
are determined by dividing the distance (2~cm) by the time shift that maximizes the correlation between 
the two normalized photodiode signals from each gun.  For the dataset presented in
Sec.~\ref{sec:results}, $V_{\rm jet}$ is generally in the range 30--45~km/s.

\subsubsection{Doppler spectroscopy}
\begin{figure}[!tb]
\centering
\includegraphics[width=2.75truein]{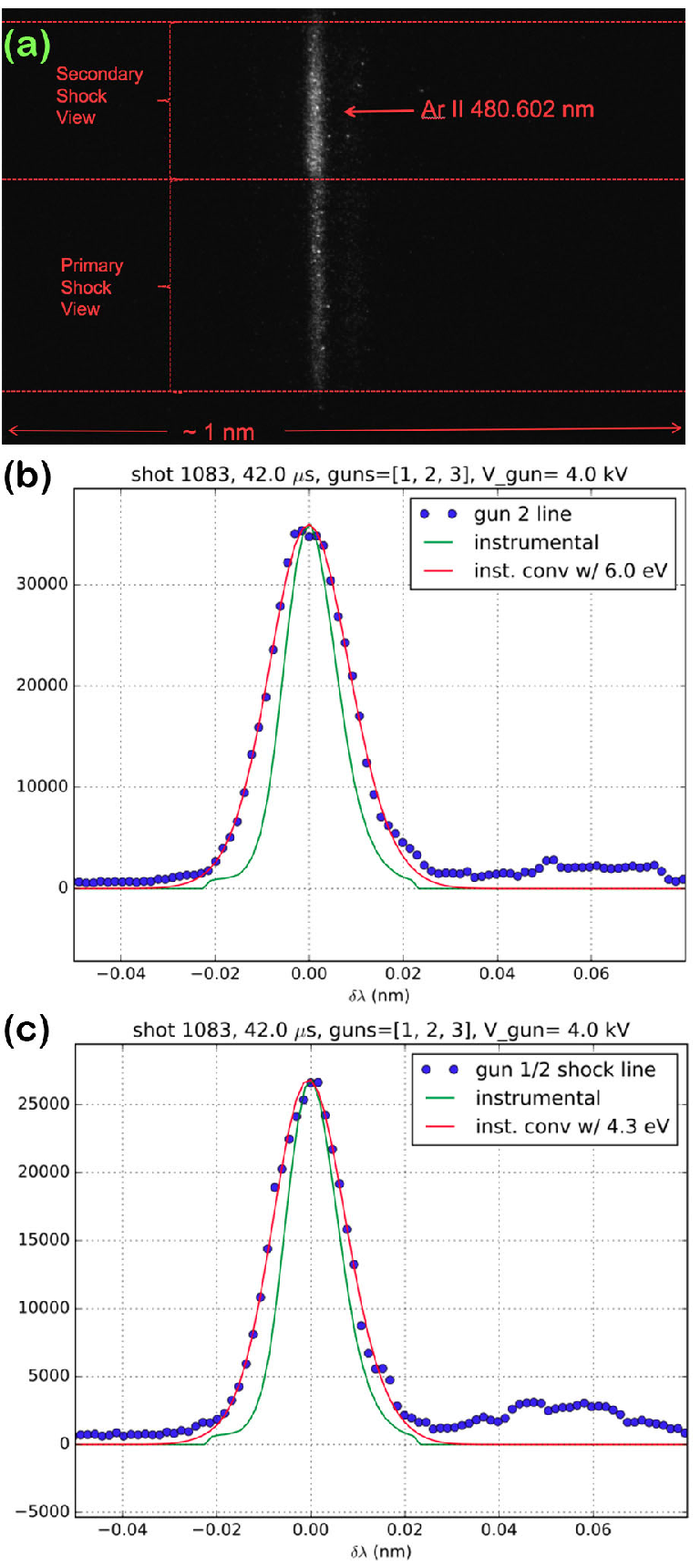}
\caption{(a)~High-resolution-spectroscopy
data (shot 1083, 1-$\mu$s exposure)
showing two views of a singly ionized argon spectral line, corresponding
to the $R=20$-cm positions shown in Fig.~\ref{fig:int_spect_setup}(b).  The vertical
and horizontal axes represent the height of the spectrometer slit and wavelength, respectively.
(b)~Vertically integrated signal (arb.)\ vs.\ wavelength (centered at 480.602~nm) for the 
secondary-shock-line view ($T_i=6.0\pm 0.13$~eV) and (c)~primary-shock-line view
($T_i=4.3\pm 0.11$~eV), from (a).  The error quoted for $T_i$ is the curve-fitting error
assuming Poisson weighting of the spectral data.}
\label{fig:spect_hi-res}
\end{figure}

This provides ion-temperature $T_i$ measurements at the shock region
between two merging plasma jets as depicted in Fig.~\ref{fig:int_spect_setup}(b).
Figure~\ref{fig:spect_hi-res}(a) shows an example of the data from the CCD detector.
Figures~\ref{fig:spect_hi-res}(b) and \ref{fig:spect_hi-res}(c) show the data, instrumental-broadening
profile, and the best fit to the data of a Gaussian convolved with the instrumental profile, for
the secondary-shock and primary-shock views, respectively.  For the cases shown
in Fig.~\ref{fig:spect_hi-res}, $T_i = 6.0 \pm 0.13$ and $4.3\pm 0.11$~eV
at $t=42$~$\mu$s and $R=20$~cm along
the secondary and primary shock lines, respectively.  Based on this time and viewing position,
$T_i=6.0$~eV is likely indicative of $T_i$ of the secondary-shock plasma.
We have also observed up to $T_i \sim 30$~eV (for argon) at the time ($t\approx 25$~$\mu$s)
and spatial position ($R\approx 50$--60~cm) of the primary shock,
but $T_i$ cools quickly (over $\sim 
10$~$\mu$s) by equilibrating with electrons (full results on ion shock heating/dynamics
for different gas species
will be reported elsewhere).  As discussed below in the survey-spectroscopy section, 
$T_e$ remains much colder throughout the primary and secondary shock-formation process.

Measurement of ion heating as an essential
property of collisional plasma shocks \cite{jaffrin64}
is an interesting study in its own right, which we are pursuing as part of a
separate project on the experimental study of plasma shocks.  Here,
our interest in shock ion heating is to provide constraining data in order
to properly assess its role in degrading the jet/liner $M
\sim C_s^{-1} \sim (T_e + T_i)^{-1/2}$.
Because $T_e$ does not increase much throughout the
jet-merging process \cite{merritt14,moser15pop} (also see survey-spectroscopy results
below) due to strong thermal and radiative loss rates, ion heating dominates the
$M$ degradation.  The latter would lead to stronger liner spreading
and is predicted to seriously degrade the ability of the liner to compress a magnetized target
plasma to reactor-relevant fusion conditions \cite{langendorf17pop}.  Ongoing research
using two-temperature ($2T$) hydrodynamic simulations (see Sec.~\ref{sec:modeling1})
is investigating the role of ion shock heating on plasma-liner formation, convergence, and
performance.

\subsubsection{Multi-chord interferometry}
\label{sec:interferometry}

\begin{figure}[!tb]
\centering
\includegraphics[width=3truein]{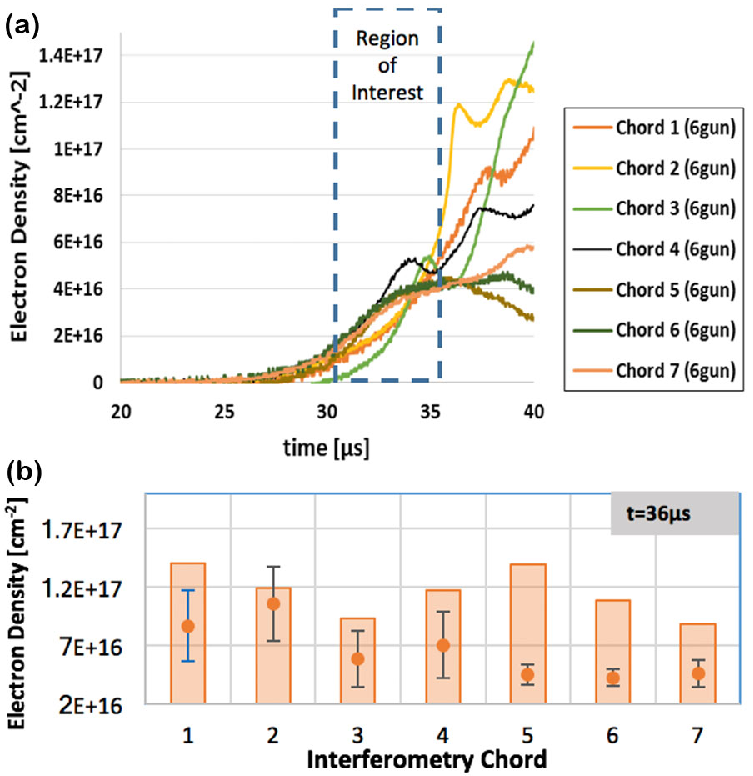}
\caption{(a)~Line-integrated electron density $\langle n_e \ell \rangle$
vs.\ time (shot 1008) from multi-chord interferometry 
[based on the setup of Fig.~\ref{fig:int_spect_setup}(a)].  ``Region of interest" refers to the approximate time 
duration when jets merge to form a liner section.  (b)~Data points are $\langle n_e \ell \rangle$ 
($t=36$~$\mu$s, averaged over shots 1019--1032) vs.\ chord number; error bars
are the standard deviation over the shot range.  The bars are synthetic data from 3D 
SPFMax hydrodynamic
simulations of the six-gun experiments (see Sec.~\ref{sec:modeling1}).}
\label{fig:int_data}
\end{figure}

Multi-chord interferometry is used to measure 
$\langle n_e \ell \rangle$ and to assess its
variation across the spatial arc of the interferometer chords shown in 
Fig.~\ref{fig:int_spect_setup}(a).  Figure~\ref{fig:int_data}(a) shows an example of $\langle n_e \ell \rangle$
for each of the seven chords vs.\ time.  Figure~\ref{fig:int_data}(b) shows a comparison
between $\langle n_e \ell \rangle$ ($t=36$~$\mu$s, averaged over shots 1019--1032) and
synthetic data from a 3D hydrodynamic simulation of the six-gun experiment (see Sec.~\ref{sec:modeling1} for a description of the simulation).  From the synthetic data, it can be seen that chords 1 and 5 are predicted to have the highest values of
$\langle n_e \ell \rangle$, consistent with those chords intersecting the position of
primary shocks [see Fig.~\ref{fig:int_spect_setup}(a)].  Similarly, chords 3 and 7 are predicted to have the
lowest values of $\langle n_e \ell \rangle$, consistent with those chords intersecting the position
of jets.

Two key, initial conclusions are drawn from the comparison between experimental
and synthetic interferometry data:  (1)~very good agreement of the
order-of-magnitude of $\langle n_e \ell \rangle$ gives
us confidence in our knowledge of the jet parameters and 
leading-order accuracy of the simulations; and (2)~poor agreement in the variation of 
$\langle n_e \ell \rangle$ vs.\ chord number is indicative
of insufficient balance (in mass and/or velocity) among jets, and thus the symmetry seen in the
synthetic data (e.g., between chords 1/5, 3/7, etc.)\ is not reproduced in the experiment.  The lack of symmetry over a wide range of time is also apparent in Fig.~\ref{fig:int_data}(a).
Numerical simulations of six-jet experiments that
incorporate unbalanced jet velocities and/or trigger times are aiding our interpretation of these data 
(see Sec.~\ref{sec:modeling1} for further discussion).
In order to improve the jet-to-jet balance, we
plan to upgrade our GVs
and have also added the ability to fine tune (through variable resistances
and inductances) the electrical characteristic of each of the six GV transmission lines.  The new
GVs (rev.~10) will provide higher precision and repeatability in the amount of mass injected.
Varying the resistance/inductance of the GV transmission lines has already allowed
us to tune the jet injected mass and velocity, as observed in the rise in chamber pressure 
after each shot, photodiode signals, and camera images.

\subsubsection{Survey spectroscopy}
\begin{figure}[!tb]
\centering
\includegraphics[width=3.2truein]{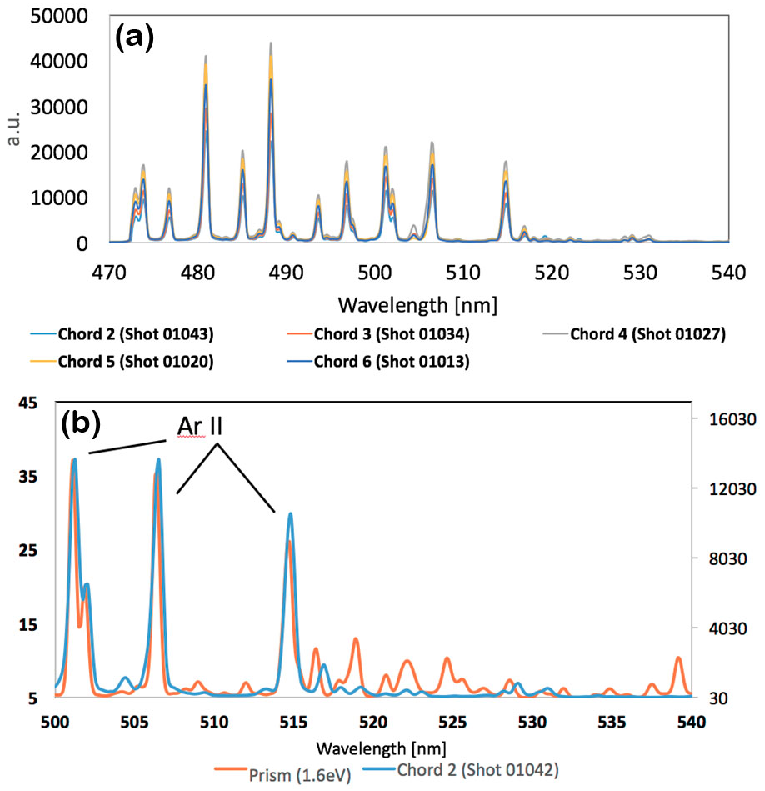}
\caption{(a)~Visible survey spectra ($t=38$~$\mu$s, 2-$\mu$s exposure)
vs.\ wavelength for views corresponding
to interferometer chords 2--6 [see Fig.~\ref{fig:int_spect_setup}(a)] for six-gun argon experiments.
(b)~Comparison of survey-spectrometer data ($t=32$~$\mu$s, 2-$\mu$s exposure)
and a PrismSPECT spectral calculation assuming
$n_e = 10^{15}$~cm$^{-3}$ and $T_e= 1.6$~eV\@.}
\label{fig:ss_data}
\end{figure}

By comparing survey-spectroscopy data with non-local-thermodynamic-equilibrium (non-LTE)
PrismSPECT \cite{macfarlane03} spectral calculations that
utilize $n_e$ values consistent with interferometry data, we are able
to place bounds on $T_e$ and $\bar{Z}$ of the observed plasma volume; this methodology was previously
described in detail \cite{hsu12pop} and applied in
multiple experimental configurations \cite{hsu12pop,merritt13, merritt14, moser15pop,
adams15pre}.  Figure~\ref{fig:ss_data}(a) shows an example
of spectra from several viewing chords, showing that there is little variation over these chords.
The chords intersect the jet-propagation axes at approximately 14.7~cm.
Figure~\ref{fig:ss_data}(b) shows an example of a
comparison between data and spectral calculations, showing
good agreement for a calculation that assumes $n_e = 10^{15}$~cm$^{-3}$
and $T_e=1.6$~eV (for which $\bar{Z}=0.99$).  On
the other hand, for calculations assuming $T_e \le 1.5$~eV
or $T_e \ge 1.9$~eV, the agreement between the spectral data and 
PrismSPECT calculations becomes dramatically worse,
implying that $1.5 < T_e < 1.9$ over a large spatial area and time range for the merging
plasma jets.

\section{Numerical modeling}
\label{sec:modeling}

In this section, we provide a brief overview of the modeling research
in support of the PLX-$\alpha$ project.
However, detailed modeling results will be reported elsewhere.

\subsection{Plasma-liner formation and convergence}
\label{sec:modeling1}

We use two 3D hydrodynamic codes to simulate jet merging (two, three, and six guns)
and spherical ($4\pi$) plasma-liner formation and convergence (18--600 guns).  The former are
in support of our ongoing experiments (studying shock dynamics between
merging jets and/or forming a section of a liner with six jets).  The latter are to guide our
preparation for planned $4\pi$ experiments over the next two years.

The two 3D codes being used are SPFMax \cite{cassibry17}, 
a smoothed-particle-hydrodynamic code, and 
FronTier \cite{samulyak07},
a hydrodynamic code with front tracking.  As part of the PLX-$\alpha$ project,
we have performed significant benchmarking of the codes against jet-merging experimental
results, and we have added physics capabilities to the codes, including Braginskii thermal
transport and viscosity \cite{braginskii65}, optically thin radiation loss, advanced equation-of-state 
(EOS) table lookup
(using custom-generated non-LTE
PROPACEOS tables \cite{macfarlane06} from Prism Computational Sciences),
and 2$T$ (i.e., separate $T_i$ and $T_e$ evolution) modeling.

SPFMax and FronTier have been used to simulate
three- and six-jet merging to guide diagnostics setup, generate synthetic data
[see Fig.~\ref{fig:int_data}(b)] for
comparisons to the diagnostic data, and aide our understanding of the comparisons.
For generating the
synthetic data in Fig.~\ref{fig:int_data}(b), we ran SPFMax using the six-gun
setup of Fig.~\ref{fig:int_spect_setup}(a) with the following initial argon-jet parameters:  
velocity of 35.8~km/s, $T_i=T_e = 2.5$~eV, ion density
$n_i = 3.17\times10^{16}$~cm$^{-3}$, diameter of 8.5~cm, length of 10~cm, and leading
edge of the jets at $R=130$~cm.  The simulation included non-LTE argon EOS, single-group
opacity, 2$T$, and ion and electron thermal conduction.  To help understand
the disagreement between the experimental and synthetic interferometry
seen in Fig.~\ref{fig:int_data}(b), we ran six-jet FronTier simulations that included
random variations among jet velocities (up to 10\%) and/or trigger times
(up to 1~$\mu$s); these results show that the symmetry of primary- and secondary-shock formation
is indeed drastically degraded compared to the case with identical jet velocities/timings
(further motivating our ongoing efforts to improve the jet-to-jet balance).
We also plan to use VISRAD \cite{macfarlane03jqsrt}
to generate synthetic spectral data to compare with spectroscopy data.

The $4\pi$ simulations (with up to $N=600$ jets)
have focused on studies of liner uniformity and ram-pressure ($\rho v^2$) evolution
as the liner forms and converges radially toward stagnation.  For $N\le 60$, the simulations
directly inform planned PLX-$\alpha$ experiments, which aim to form $4\pi$ imploding liners
with 36--60 jets and $\sim 100$--150~kJ of total liner kinetic energy, and predict peak $\rho v^2 \sim 50$~kbar. 
For $N>60$, the simulations are assessing the jet and uniformity requirements 
to achieve fusion-relevant conditions, requiring
total liner kinetic energy of $\gtrsim 30$~MJ and peak $\rho v^2\gtrsim 150$~Mbar 
\cite{knapp14,langendorf17pop}.

\subsection{Plasma-liner compression of a magnetized target}
\label{sec:modeling2}

A separate modeling task is studying plasma-liner compression of a magnetized target in
1D and 2D, using the multi-fluid magnetohydrodynamic (MHD) code USim \cite{beckwith15}.
A major purpose of this task is to identify and optimize PJMIF configurations with energy gain,
guided by the results of \cite{knapp14,langendorf17pop}.
New physics capabilities have also been added to USim
as part of the PLX-$\alpha$ project, including optically thin radiation loss, non-LTE tabular EOS
table lookup (also using PROPACEOS tables \cite{macfarlane06}), Braginskii viscosity 
\cite{braginskii65}, and $\alpha$-particle energy deposition based on the model implemented
in \cite{langendorf17pop}.  A second important purpose of this task is to assess and
understand the degradation of energy gain in going from 1D to 2D, and to examine the further
degradation of energy gain when nonuniformities (based on
the work described in Sec.~\ref{sec:modeling1})
are imposed on the liner/target interface, which exacerbate Rayleigh-Taylor
instabilities at the stagnating interface.  Indeed,
one specific objective of this task is to set requirements on the liner uniformity for PJMIF to
remain viable as a fusion-energy concept.

\section{Summary and plans}
\label{sec:summary}

We have described an experiment to form and characterize a section of a spherically
imploding plasma liner, as a development step toward PJMIF
\cite{thio99,hsu12ieee}.
This work is the first phase of the ARPA-E-sponsored
PLX-$\alpha$ project, which aims to culminate with the formation
and study of a spherically imploding plasma liner (as a standoff MIF compression driver)
formed by merging 36--60 supersonic plasma jets.  The latter are launched by newly 
developed coaxial plasma guns fabricated by HyperV Technologies Corp.\ (now owned
by new fusion startup HyperJet Fusion Corporation).

This paper reports key early results from the PLX-$\alpha$ project:  (1)~design
and operation of the new HyperV coaxial plasma guns and characterization
of plasma-jet parameters, (2)~completion of
the PLX facility/diagnostic upgrades for six-gun experiments, (3)~successful operation of 
up to six plasma guns and key diagnostics (photodiode arrays, fast imaging
cameras, survey spectroscopy,
hi-resolution spectroscopy, and multi-chord interferometry), and (4)~diagnostic results
showing that the key potentially deleterious physics issues associated with jet merging
(e.g., ion shock heating leading to Mach-number-degradation and the seeding of nonuniformities
that may exacerbate deceleration-phase instabilities of the liner/target interface
in future target-compression experiments)
can now be studied in a serious manner; these 
studies are ongoing, and further details and conclusions will be reported elsewhere.

Ongoing three- and six-jet
experiments are providing comprehensive datasets on ion shock heating for 
several plasma-jet species (H, He, N, Ne, Ar, Kr, Xe)
and two different jet-merging angles.  Upon installation
of improved gas valves, which are
expected to improve the jet-to-jet balance, we will then focus on obtaining
more interferometry data to better assess the uniformity of the liner section formed by merging
six jets [i.e., the setup shown in Fig.~\ref{fig:int_spect_setup}(a)].  Finally, we are
incorporating a number of engineering improvements to the coaxial plasma guns to maximize
both the assembly/operational efficiency and quality/quantity of experimental data for 
planned experiments with
36--60 guns to form $4\pi$ spherically imploding plasma liners (over the next two years).
These experiments will provide critical data on the magnitude
and evolution of liner nonuniformity and
ram pressure during radial convergence, to allow continued technical assessment 
and development of the PJMIF concept.

\section*{Acknowledgments}

The authors thank R.~Aragonez, R.~Martinez (dec.), J.~Vaughan, and M.~Luna for valuable
technical contributions,
and Dr.~G.~Wurden for loaning numerous items of diagnostic hardware.
Section~III of this paper summarizes
the invited talk by S.~Hsu and posters by S.~Langendorf and K.~Yates that were presented at
the International Conference on Plasma Science (ICOPS) in Atlantic City, NJ, May 22--25, 2017.
This work was supported by the Advanced Research Projects Agency--Energy (ARPA-E) of the
U.S. Department of Energy (DOE)\@.  We also
acknowledge the DOE Office of Fusion Energy Sciences for
sponsoring PLX construction and research (2009--2012) and plasma-gun development
by HyperV Technologies Corp.\ (2005--2012).

\ifCLASSOPTIONcaptionsoff
  \newpage
\fi

\end{document}